\documentclass[fleqn,usenatbib]{mnras}
\usepackage{gensymb}
\usepackage[T1]{fontenc}
\usepackage{ae,aecompl}
\usepackage{graphicx}	
\usepackage{mathtools}	
\usepackage{amssymb}	
\usepackage{xspace}	
\usepackage[dvipsnames]{xcolor}

\newcommand{\package}[1]{\textsl{#1}}

\newcommand{\msun}{\textrm{M}_\odot}
\newcommand{\kpc}{\textrm{kpc}}
\newcommand{\pone}{\phi_1}
\newcommand{\ptwo}{\phi_2}

\newcommand{\cugi}{C_{ugi}}
\newcommand{\dtt}{|\ptwo - f(\pone)|}
\newcommand{\dCN}{\ensuremath{\delta{\rm CN}}\xspace}
\newcommand{\dCH}{\ensuremath{\delta{\rm CH}}\xspace}

\newcommand{\kms}{\ensuremath{\textrm{km}~\textrm{s}^{-1}}}

\newcommand{\feh}{\ensuremath{[\textrm{Fe} / \textrm{H}]}}
\newcommand{\noop}[1]{}

\title[Evidence for C and Mg variations in GD1]{Evidence for C and Mg variations in the GD-1 stellar stream}

\author[Balbinot, Cabrera-Ziri, and Lardo]{
Eduardo Balbinot$^{1}$\thanks{E-mail: balbinot@astro.rug.nl},
Ivan Cabrera-Ziri$^{2}$, and 
Carmela Lardo$^{3}$
\\
$^{1}$Kapteyn Astronomical Institute, University of Groningen, Landleven 12,
NL-9747 AD Groningen, the Netherlands \\
$^{2}$Astronomisches Rechen-Institut, Zentrum f\"ur Astronomie der
Universit\"at Heidelberg, M\"onchhofstra{\ss}e 12-14, D-69120 Heidelberg,
Germany \\
$^{3}$Dipartimento di Fisica e Astronomia, Universit\`a degli Studi di Bologna,
Via Gobetti 93/2, 40129, Bologna, Italy\\
}

\date{in original form \today}

\pubyear{2020}

\begin{document}
\label{firstpage}
\pagerange{\pageref{firstpage}--\pageref{lastpage}}
\maketitle

\begin{abstract}
Dynamically cold stellar streams are the relics left over from globular cluster
dissolution. These relics offer a unique insight into a now fully disrupted
population of ancient clusters in our Galaxy. Using a combination of Gaia eDR3
proper motions, optical and near-UV colours we select a sample of likely Red
Giant Branch stars from the GD-1 stream for medium-low resolution spectroscopic
follow-up. Based on radial velocity and metallicity, we are able to find 14 new
members of GD-1, 5 of which are associated with the \emph{spur} and
\emph{blob/cocoon} off-stream features. We measured C-abundances to probe for
abundance variations known to exist in globular clusters. These variations
are expected to manifest in a subtle way in globular clusters with such low
masses ($\sim10^4 {\rm ~\msun}$) and metallicities (${\rm[Fe/H]}\sim-2.1 {\rm
~dex}$).  We find that the C-abundances of the stars in our sample display a
small but significant ($ 3\sigma$ level) spread. Furthermore, we find
$\sim 3\sigma$ variation in Mg-abundances among the stars in our sample that
have been observed by APOGEE. These abundance patterns match the ones found in
Galactic globular clusters of similar metallicity. Our results suggest that
GD-1 represents another fully disrupted low mass globular cluster where
light-element abundance spreads have been found. 
\end{abstract}


\section{Introduction}

Among the myriad of substructure present in the Milky Way (MW) halo, stellar
streams stand out as being spatial and kinematic cohesive structures which in
some cases span hundreds of degrees on the sky.  In recent years, dozens of new
streams have been discovered \citep[e.g.][]{Bernard:2016, Balbinot:2016,
Malhan:2018, Shipp:2018a, Ibata:2019} and the community has exploited their
properties to refine our knowledge about the mass and shape of the Galaxy
as well as its accretion history \citep[e.g.][]{Kuepper:2015,Bonaca:2018,
Massari:2019}.

Streams are classified as \emph{hot} or \emph{cold}, based on the dynamical
temperature (i.e. velocity dispersion) of their progenitor.  The low velocity
dispersion of their progenitor makes \emph{cold} stellar streams intrinsically
thin ($\lessapprox 100$ pc), pointing to globular clusters (GCs) as their
precursors. Palomar 5 offers a spectacular example of a \emph{cold} stream with
tidal tails radiating from a still bound cluster
\citep[e.g.][]{Odenkirchen:2001}. However, in our Galaxy, \emph{cold} streams
like Palomar 5 are rare.

In fact, in the Milky Way most cold streams are progenitor-less
\citep[e.g.][]{Shipp:2018a}, this means that their GC progenitors are now
completely dissolved. Yet, the vast majority of GCs in the Galaxy show no signs of
having streams in formation \citep[e.g.][]{ Kuzma:2018,Sollima:2020}.

It is natural to ask the origin of such disparity between the numbers of GC
with streams and progenitor-less cold streams. It is entirely possible that the
GCs that gave rise to the population of progenitor-less streams were sampled
from a distribution of long-gone GCs that differ from present-day MW GCs. One
way of producing these fast-dissolving GCs is by retaining a large fraction of
stellar-mass black holes, as demonstrated by \citet{Gieles:2021}. The authors
propose two natural pathways, a low initial density or a flatter initial mass
function; as well as nurture pathways, such as tidal heating and mass stripping
during accretion. In this context, progenitor-less streams could be probing a
population of intrinsically distinct GCs than the present-day ones in the MW.
Alternatively, their progenitors may have formed in a host galaxy that is now
accreted into the MW. In any of these cases, studying their remnants (stellar
streams) may provide insight into the population of GCs no longer available in our
Galaxy.

Although historically described as simple stellar populations, in reality, GCs
are far from it. The abundances of light elements like e.g. He, C, N, O are
known to change from star to star within a cluster
\citep[e.g.][]{Charbonnel16,Gratton19}. These multiple stellar populations
(MPs) seem ubiquitous in high mass GCs ($\gtrsim10^5 \msun$) with ages
$\gtrsim2$~Gyr but its presence has not been detected in low mass systems like
Galactic open clusters \citep[e.g.][]{Carrera13,MacLean15} including older ones
like NGC 6791 \citep[e.g.][]{Bragaglia14,Cunha15}. The current interpretation
for this is that the mechanism responsible for these abundance variations
require high-mass and/or density and/or redshift to operate \citep[see][and
references therein]{BL18}. Unfortunately, very little is known about the
presence/absence of MPs in low mass old ($\sim10$~Gyr) stellar clusters. The
reason for this is that such systems are very rare as their chances to get
disrupted/dissolved are very high in these timescales. 

There are approximately ten globular clusters with masses $\lesssim
10^4~\msun$ \citep[see][]{Baumgardt18,Vasiliev:2021}. Some of them have been
the focus of studies searching for MPs, for example, NGC 6535 and ESO452-SC11 at
$\sim3\times 10^3\msun$ and $\sim 8\times 10^3\msun$, respectively
\citep[see][]{Baumgardt18,Vasiliev:2021}, show some evidence for abundance
variations \citep[e.g.][]{Bragaglia:2017, Simpson:2017}. On the other hand, in
other low mass globular clusters like E3 ($\sim 3\times 10^3~\msun$) and Rup
106 ($\sim 3\times 10^4~\msun$) the presence of MPs has not been detected
\citep[e.g.][]{Monaco18,Dotter18}. However, the slopes of their present day
mass functions of these clusters are very shallow/evolved with respect to a
Kroupa mass function \citep[see][]{Baumgardt18,Vasiliev:2021}, as is expected
for clusters that have suffered a strong dynamical evolution (mass loss).
Understanding the behaviour of MPs in globular clusters in the low end of the
globular cluster mass distribution would provide valuable insights into the
mechanism responsible for their origin (which remains not very well
understood).

The initial mass of cold stellar streams can be more reliably constrained than
for MW GCs. This is because the orbit of streams can be accurately measured,
and the total number of ejected stars can be estimated from the stream itself,
while for GCs, these ejected stars may be present, but too low-mass to be
detected. Thus, stellar streams provide a more robust candidate for studying
MPs in a truly low-mass regime.

Additionally, the MPs are known to be spatially segregated in many GCs
\citep[e.g.][]{Lardo:11,Larsen:15,Dalessandro:19,Leitinger22} and simulations
show that the two populations can remain unmixed during most of a GC lifetime
\citep[e.g.][]{HenaultBrunet:2015}.  If that is the case, the outer population
will escape first, leaving this spatial segregation imprinted on the
distribution of stream stars.  This could be exploited to pinpoint the position
of the fully dissolved progenitor, ultimately leading to better dynamical
models for stream formation.

The GD-1 stream was discovered using Sloan Digital Sky Survey
\citep[SDSS;][]{Eisenstein:2011} by \citet{Grillmair:2006} and in recent years
attracted much attention thanks to the Gaia DR2 \citep{GaiaCollaboration:2018}
which allowed it to be studied in unprecedented detail
\citep{PriceWhelan:2018}. Using collisional N-Body simulations
\citet{Webb:2019} model the dissolution process of GD-1, and conclude its
progenitor must have been a very low-mass GC (a few $10^4 ~\msun$), that
dissolved at most in the last 3 Gyrs. Their conclusions would place the
progenitor of GD-1 in the low-mass edge of the mass distribution of MW GCs and
that the GD-1 progenitor would fully dissolve in a fraction of a Hubble time,
possibly pointing to an extragalactic origin or atypically GC
evolution/formation. The dynamical mass estimate is supported by
observational constraints on the total luminous mass of GD-1 by
\citet{deBoer:2020}, who report a mass of $1.58 \pm 0.07 \times 10^4~\msun$.
The GD-1 location in the outer halo provides a much better case for a truly
initially low-mass GC counterpart. We also highlight the recent discovery of
evidence for mutilple stellar populations in the Phoenix stream
\citep{Balbinot:2016, Wan:2020}, with similarly low mass to GD-1, however with
remarkably low metallicity. The oddities of stellar stream, i.e. their low
mass and metallicity, seem to indicate that these objects belong to a distinct
class of GCs, perhaps of extra-galactic origins, as discussed previously.

The peculiar characteristics of the GD-1 progenitor place it in a mass-age
regime that is not accessible in the present population of GCs. Thus, it is
interesting to investigate other properties unique to GCs. As mentioned above,
some hypotheses propose that the conditions similar to the ones found in the
discs of high redshift galaxies (i.e. large gas fractions, high gas densities,
high turbulent speeds) enable the mechanisms responsible for producing the MPs
characteristic of GCs
\cite[e.g.][]{DErcole:2016,Elmegreen:2017,Gieles:2018,Johnson:2019}.  The
implication is that stars that form in low-density environments should not host
MPs.

In section 2 we present the data used to select GD-1 candidate members.  In
section 3 we follow-up on these candidates and obtain spectra to derive radial
velocities, metallicities, and abundance information. In section 4 we discuss
other chemical signatures found in GD-1. And in section 5 we discuss our
results and their implication.

\section{Data}

\begin{figure}
    \centering
    \includegraphics[width=0.35\textwidth]{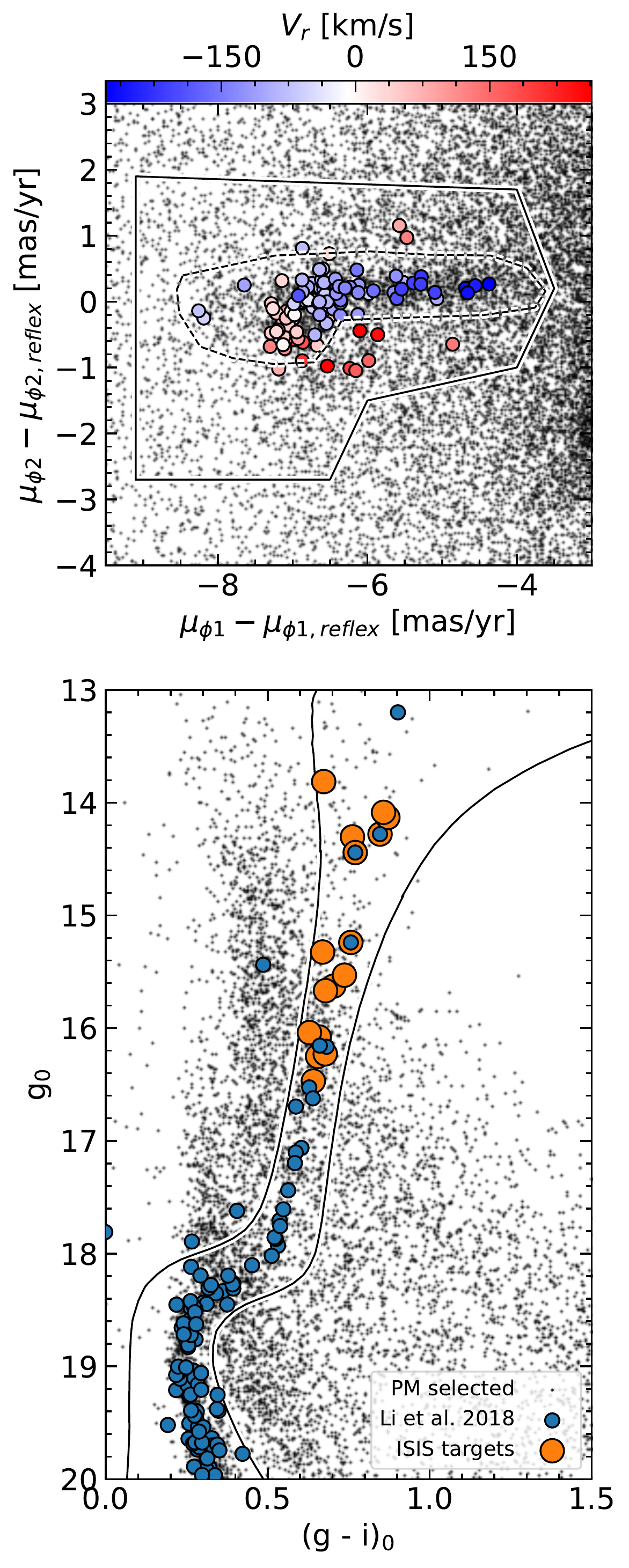}
    \caption{\emph{Top panel:} PM distribution of CMD selected stars (see panel
        below) and with $\dtt<2\degree$. The \citet{Li:2018} sample is overlaid
        (blue). The colour scale shows the line-of-sight velocity. \emph{Bottom
        panel:} $(g-i)$ \emph{vs} $g$ CMD. The solid lines show a selection
        based on a PADOVA isochrone of $\log_{10} {\rm age}/{\rm yr} = 10.03$
        and $\feh = -2.2$.  We also mark targets that were selected for
        spectroscopic follow-up (see Sec. \ref{sec:spec}).}
    \label{fig:selection}
\end{figure}

To select GD-1 members, we use the Gaia Data Early Release 3
\citep[eDR3;][]{GaiaCollaboration:2020b}, where we cross-match with LAMOST DR6
\footnote{\url{http://dr6.lamost.org/v2.0/}} (including the subsample from
\citealt{Li:2018}), SDSS DR13 photometry \citep{Eisenstein:2011}, Pan-STARSS
DR1 \citep{Chambers:2016}, and 2MASS \citep{Skrutskie:2006}. We prefer official
Gaia cross-matches (Pan-STARRS and 2MASS), however when not available we use
the Whole Sky Database (see acknowledgments). For smaller catalogues extracted
from literature, we match by Gaia eDR3 \texttt{source\_id} or through a
positional match using STILTS \citep{Taylor:06}. We also note that LAMOST has a
more recent DR7, however for comparing our membership selection
with literature we adopt DR6. Later in the paper, we address the newer data.

For the astrometric catalogue of Gaia eDR3, we adopt the following quality cut:
\texttt{RUWE} $< 1.4$ \texttt{\&} \texttt{visibility\_periods\_used} $> 3$.  We
remove some foreground stars using a parallax $< 1$ selection.  We limit our
catalogue to the reported region occupied by GD-1 as reported by
\citet{DeBoer:2018}. All magnitudes in our sample were corrected for extinction
using \citet{Schlegel:1998} maps and \citet{Cardelli:1989} extinction law, with
$\mathrm{R_V} = 3.1$.

The LAMOST/SDSS sample from \citet{Li:2018} contains multiple observations of some
stars. For these duplicates, we compute the average of their line-of-sight
velocities and metallicities (when available). Uncertainties were propagated
accordingly.

For our analysis, it is useful to define a rotated coordinate system ($\pone,
\ptwo$)  where the equator is approximately aligned with the GD-1. This system
is defined via a rotation matrix given by \citet{Koposov:2010}. We adopt the
distances from this same work to correct the proper motions (PMs) for the Solar
reflex motion. The \citet{Schonrich:2010} reflex motion is used, while the
Sun's distance to the Galactic centre is assumed to be 8.3 $\kpc$
\citep{Gillenssen:2009}. We also use the best-fit 3D position of the stream to
bring the stars to a common distance of 8.3 $\kpc$, which is the stream
distance at its midpoint. Throughout this paper, we use only distance
normalised and de-reddened magnitudes. We also prefer Pan-STARRS magnitudes
when not using the $u$-band.

\subsection{Membership selection}
\label{sec:memb}

Following a similar procedure as outlined by \citet{PriceWhelan:2018} we define
a polygon cut in PM space. This polygon was constructed based on a sample of
main-sequence stars selected from a CMD cut and distance to the stream track
\citep[as defined by][]{DeBoer:2018} less than 2\degree~ (i.e. $\dtt <
2\degree$; where $f(\pone)$ is the interpolated stream track).  To
define a more robust colour-magnitude selection, that includes Red Giant Branch
(RGB) stars, we use the PM selected sample and the spectroscopically confirmed
sample to define a narrow colour-magnitude cut. When compared to the PM
selection from \citet{PriceWhelan:2018}, or selection is more restrictive,
mainly due to RGB stars being brighter, thus having lower astrometric
uncertainties. We also note that some spectroscopically confirmed members are
not included in our PM selection, these members are typically close to the lower
limit in $\pone$, which was not favoured in our selection due to observability
constraints (see Sec. \ref{sec:spec}).

In Figure \ref{fig:selection} we show a summary of our selection process. In
the top panel, the PM selection polygon is shown, while the sample from
\citet{Li:2018} is over-plotted. The colour scale indicate the radial velocity
value. On the bottom panel, the colour-magnitude diagram (CMD) for the
PM-selected sample is shown. To define a CMD selection around likely GD-1
members we use a PARSEC isochrone \citep{Bressan:2012}. Our best-fit isochrone
is defined as the one that well represents the main-sequence, and is able to
reproduce the observed RGB members from \citet{Li:2018}. It is out of the scope
of this paper to produce a robust isochrone fit, we are interested only in
defining a region of the CMD that contains likely GD-1 member stars. For this
we defined a range of colours around the best-fit isochrone based on the
photometric uncertainties while allowing for a wider range in colours at
bright magnitudes in order to include regions of the CMD where confirmed GD-1
members occupy.  When necessary, we also defined colour offsets to our mask to
encompass all spectroscopic members of GD-1, except for two likely
Blue Stragglers in the \citet{Li:2018} sample in the same Figure we also define
a new PM selection based on the spectroscopic members, as discussed above.

In the literature, combinations of filters have often been used to pick up the
variations in specific spectral features characteristic of the MPs phenomena
\citep[e.g.][]{Marino08,Lardo11,Monelli13}. Here we exploit the $\cugi$ colour
index, which is sensitive to CN molecular bands around 385 nm originating from
cold stellar atmospheres. This index is defined as:

\begin{align}
\cugi &= (u-g) - (g-i) \\
\sigma_{ugi} &= \sqrt{\sigma_u^2 + 2 \sigma_g^2 + \sigma_i^2}
\end{align}
where $\sigma_{ugi}$ is the photometric uncertainty propagated from the 
uncertainties in $u$, $g$ and $i$. 

\begin{figure}
    \centering
    \includegraphics[width=0.45\textwidth]{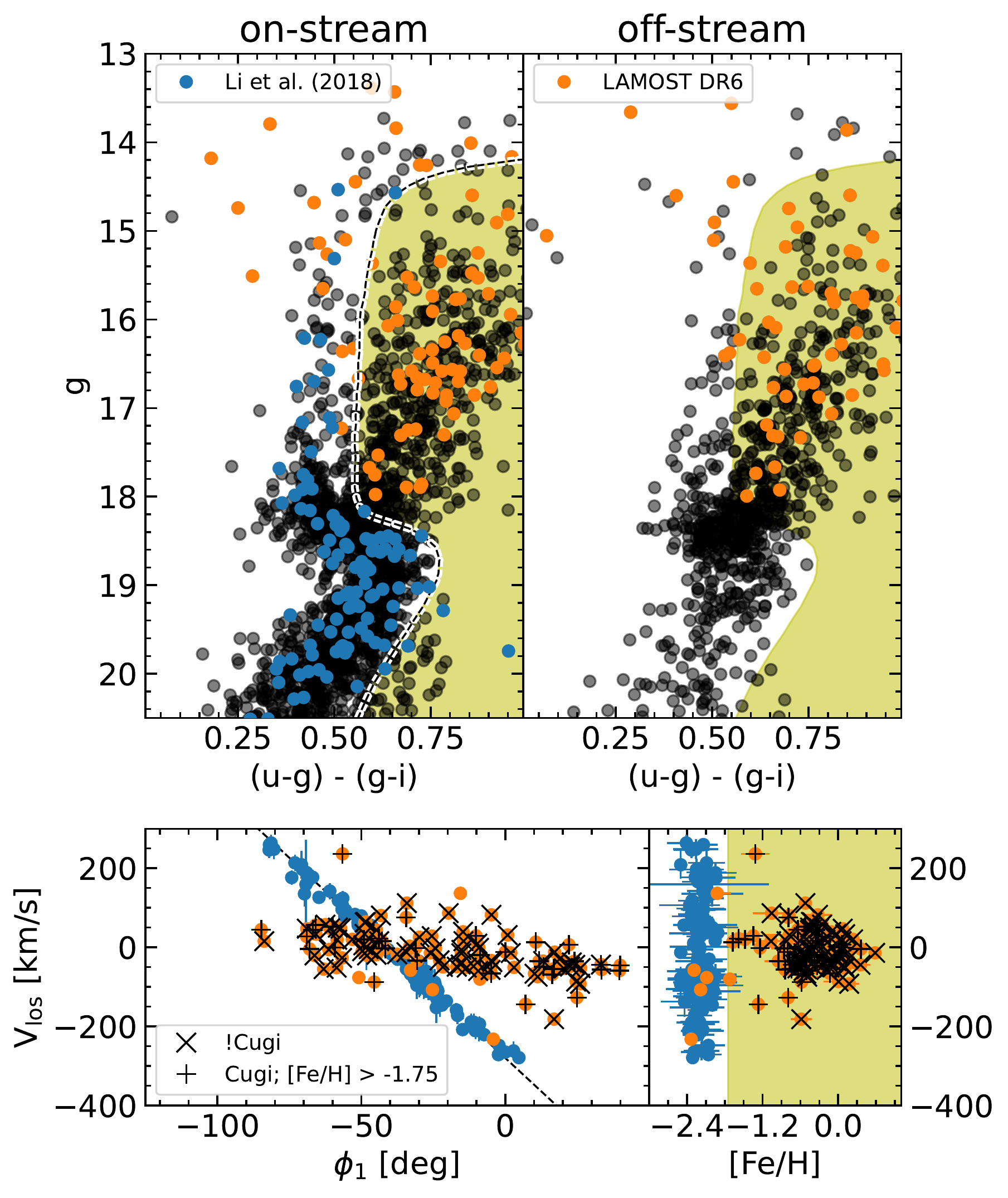}
    \caption{\emph{Top left:} $\cugi$ CMD of the narrow-PM and optical-CMD
        selected sample. We highlight the \citet{Li:2018} (blue) and LAMOST DR6
        samples (orange). We overlay the same isochrone used in Figure
        \ref{fig:selection}, shifted in colour (dashed line) to
        define a selection of potential members. The yellow shaded region marks
        the CMD region likely to contain only field stars. \emph{Top right:}
        same as previous panel, but for \emph{off-stream} stars ($|\ptwo -
        f(\pone) - 4\degree | < 2\degree)$.  \emph{Bottom left:} $v_r$ as a function of
        $\pone$, the dashed line is the best-fit orbit from
        \citet{Koposov:2010}. We overplot $\times$ on stars that are likely to
        be a contaminant based on their position in the $\cugi$ CMD. Stars marked
        with $+$ signs are removed from the sample since their LAMOST DR6
        metallicities do not comply with the bulk of GD-1 stars (\emph{bottom right
        panel}).}
    \label{fig:cugivr}
\end{figure}

Besides being an indicator for MPs, this colour index is also highly sensitive
to metallicity. In Figure \ref{fig:cugivr} we show the $\cugi$ CMD for stars
selected based on PM (narrow selection) and optical CMD (see Figure
\ref{fig:selection}). We show the CMD for an \emph{on-stream} region (top left
panel), defined as $|\ptwo - f(\pone)|<2\degree$, and an \emph{off-stream}
region (top right panel), defined as $|\ptwo - f(\pone) - 4\degree|<2\degree$.
The \emph{off-stream} shows the CMD locus occupied by MW field stars that
comply with both our PM and CMD cuts. Based on the best-fit isochrone, we
define a region where contaminants are likely to be found (highlighted in
yellow). Notice that due to the uncertainty of the isochrone fit, we allow for
a broader range in colour at brighter magnitudes. To confirm that these stars
are contaminants we show their radial velocity distribution as a function of
$\pone$ (i.e. along the stream).  When compared to the best-fit orbit from
\citet{Koposov:2010} (dashed line) we see that many of the stars have
velocities incompatible with the stream. A few stars overlap with the expected
velocity trend, but on closer inspection, their metallicity is inconsistent
with the bulk of the GD-1 stars (bottom right panel). We find that the $\cugi$
selection is able to remove most of the contaminants, and the few ones left
either have incompatible velocities and/or metallicities. We also note the
existence of a few LAMOST DR6 stars that could be GD-1 members and that are not
in the \citet{Li:2018} sample. We follow-up on these potential members in the more
recent \mbox{LAMOST} DR7 and find that several of these stars have multiple epoch
observations. We compute their averaged spectra, metallicities, and velocities
from the LAMOST database. We select GD-1 members with radial velocities
consistent ($<$ 3$\sigma_{V_{los}}$) with the stream orbit and [Fe/H] $<$
-1.75. We use this data in conjunction with the newly obtained spectra
presented in the following section.

\section{Spectroscopic follow-up}
\label{sec:spec}

\begin{figure}
    \centering
    \includegraphics[width=0.5\textwidth]{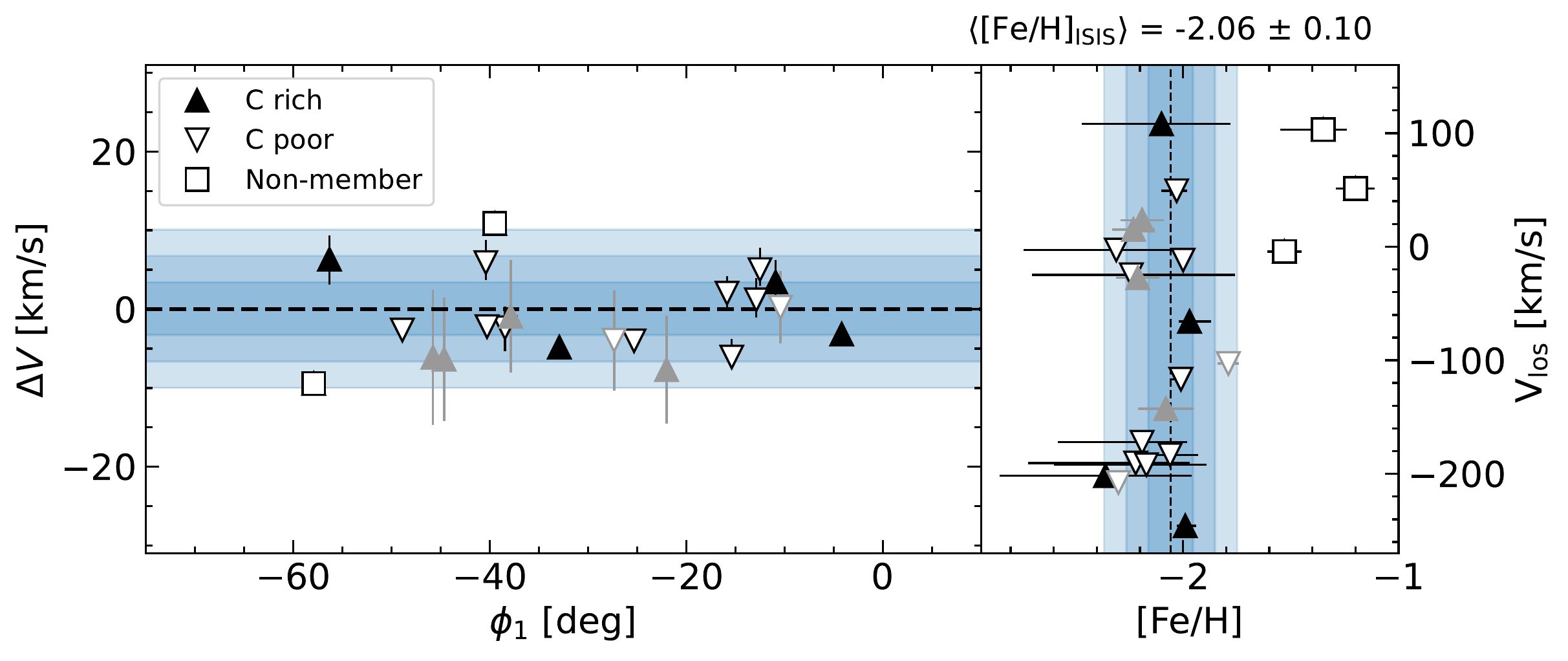}
    \caption{ISIS red-arm radial velocity \emph{vs} $\pone$ (\emph{left}) and
        metallicity (\emph{right}).  Members are marked with up/down triangles
        depending on their measured C-abundance (see Section
        \ref{sec:abundances}), non-members are shown as squares. On the
        \emph{left} the velocities have been subtracted by the best-fit
        \citet{Koposov:2010} orbit. Both panels show the mean (dashed line) as
        regions 1, 2, and 3 $\sigma$ away from it (shaded blue tones), computed
        using a uncertainty-weighted average. We note that one of the
        high-metallicity non-members lie outside of the velocity range in the
        \emph{left panel.} In both panels we show LAMOST DR7 stars in grey,
        following the same notation as before. These were not used to compute
        the confidence intervals and average metallicity.
        } 
    \label{fig:isismemb}
\end{figure}

\begin{figure*}
    \centering
    \includegraphics[width=0.95\textwidth]{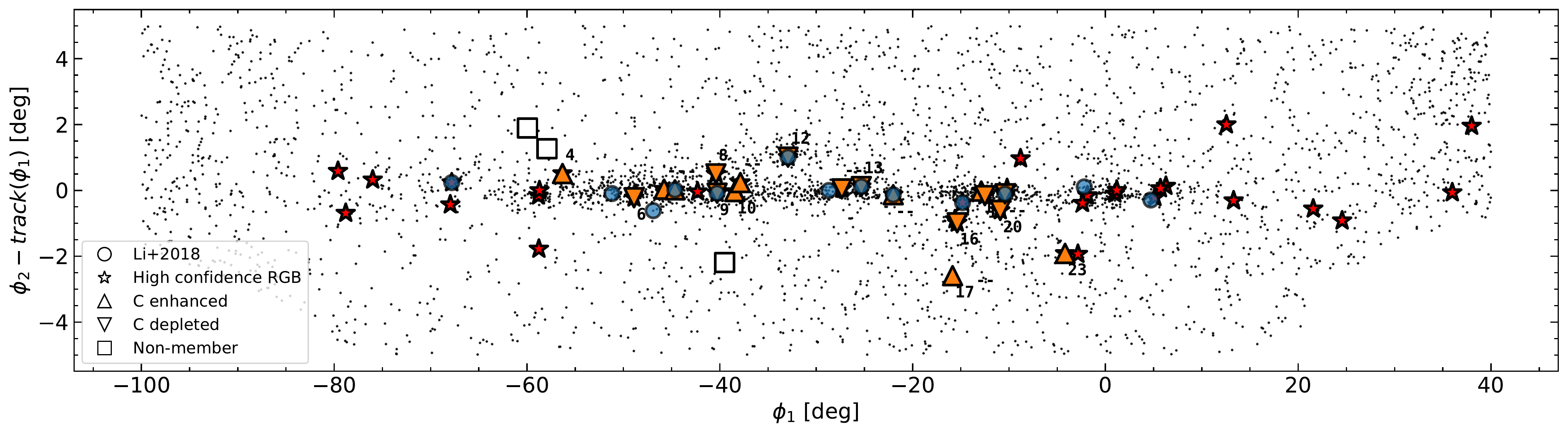}
    \caption{Spatial distributions of CMD (optical and near UV) and PM (narrow)
        selected stars (black). High probability RGB stars in the magnitude
        range of $14 < g < 17$ are shown as red stars.  We split our
        spectroscopic sample into three groups, C rich/poor (see Section
        \ref{sec:abundances}), and non-members (see Figure \ref{fig:isismemb}).
        Stars in our ISIS spectroscopic sample are labelled by their ID (see Table
        \ref{tab:summary}). The \citet{Li:2018} sample is shown as blue circles.} 
        \label{fig:spatial}
\end{figure*}

\begin{table*}
\scriptsize
\caption{IDs, eDR3 \texttt{source\_id}, rotated coordinates, G-magnitude, blue-arm atmospheric parameters; red-arm radial velocities and metallicity measurements of our targets. Stars in common with \citet{Li:2018} and APOGEE DR16 \citep{Ahumada20} have their \texttt{source\_id} marked with an $^*$ and $^\dagger$, respectively. 
We also list the effective temperature (T$_{eff, \mathrm{phot}}$) and surface gravity ($\log g_{\mathrm{phot}})$ derived from photometry (see Sec. \ref{sec:spec}). For LAMOST DR7 stars, we list the metallicity and velocity measured by their pipeline, taking their average when multi-epoch data is available. 
}
\label{tab:summary}
\begin{tabular}{lcrrcllllrl}
\hline
ID    & \texttt{source\_id}              & $\pone$     & $\ptwo$    & G       & $T_{eff, blue}$     & $\log g_{blue}$        & T$_{eff, \mathrm{phot}}$ & $\log g_{\mathrm{phot}}$ & $V_{los}$                 & [Fe/H]                  \\
-     & -                                & deg         & deg        & mag     & K                   & dex                    & K                      & dex                    & km/s                      & dex                     \\
\hline 
      &                                  &             &            &         &                     & ISIS Members           &                        &                        &                           &                         \\
\hline
$4$   & $686849456285987840^{\dagger}$   & $-56.31766$ & $0.15197$  & $15.43$ & $5240^{+52}_{-65}$  & $2.35^{+0.15}_{-0.29}$ & 5365 $\pm$ 38          & 2.71 $\pm$ 0.04        & $98.39^{+3.04}_{-3.21}$   & $-2.10^{+0.32}_{-0.37}$ \\
 $6$  & $696138061798355840$             & $-48.90048$ & $-0.26598$ & $14.55$ & $5138^{+15}_{-77}$  & $1.82^{+0.07}_{-0.65}$ & 5181 $\pm$ 41          & 2.32 $\pm$ 0.05        & $49.87^{+0.90}_{-1.08}$   & $-2.03^{+0.05}_{-0.07}$ \\
 $8$  & $796283127046157312$             & $-40.40242$ & $0.71651$  & $15.29$ & $5355^{+38}_{-61}$  & $2.43^{+0.12}_{-0.26}$ & 5299 $\pm$ 79          & 2.65 $\pm$ 0.06        & $-13.31^{+2.86}_{-2.26}$  & $-2.31^{+0.33}_{-0.43}$ \\
 $9$  & $796217426927895168^{* \dagger}$ & $-40.28003$ & $0.13156$  & $13.42$ & $5299^{+5}_{-5}$    & $2.00^{+0.03}_{-0.01}$ & 4918 $\pm$ 12          & 1.86 $\pm$ 0.06        & $-11.75^{+0.54}_{-0.61}$  & $-2.00^{+0.02}_{-0.01}$ \\
 $10$ & $796779998924250624$             & $-38.45281$ & $0.17051$  & $15.56$ & $5255^{+58}_{-68}$  & $2.03^{+0.30}_{-0.36}$ & 5395 $\pm$ 46          & 2.80 $\pm$ 0.05        & $-23.28^{+2.71}_{-3.02}$  & $-2.24^{+0.48}_{-0.46}$ \\
 $12$ & $805475834527735168^*$           & $-32.92161$ & $1.13310$  & $13.26$ & $5036^{+6}_{-3}$    & $1.00^{+0.25}_{-0.05}$ & 4838 $\pm$ 58          & 1.64 $\pm$ 0.05        & $-64.71^{+1.08}_{-0.83}$  & $-1.97^{+0.10}_{-0.05}$ \\
 $13$ & $832080270707721088^*$           & $-25.31829$ & $0.25339$  & $14.45$ & $5051^{+73}_{-3}$   & $1.03^{+0.90}_{-0.93}$ & 5123 $\pm$ 62          & 2.18 $\pm$ 0.05        & $-114.73^{+0.99}_{-1.22}$ & $-2.01^{+0.03}_{-0.05}$ \\
 $16$ & $840073411003209216$             & $-15.37631$ & $-1.02189$ & $15.13$ & $5164^{+62}_{-101}$ & $1.62^{+0.27}_{-0.63}$ & 5216 $\pm$ 46          & 2.44 $\pm$ 0.06        & $-175.37^{+2.30}_{-1.53}$ & $-2.06^{+0.13}_{-0.18}$ \\
 $17$ & $790916754387404928$             & $-15.85484$ & $-2.62079$ & $14.78$ & $5370^{+35}_{-43}$  & $2.50^{+0.19}_{-0.19}$ & 5479 $\pm$ 56          & 2.24 $\pm$ 0.06        & $-169.52^{+2.12}_{-2.16}$ & $-2.19^{+0.21}_{-0.39}$ \\
 $18$ & $840818192691653504$             & $-12.87469$ & $-0.09236$ & $15.61$ & $5200^{+57}_{-36}$  & $1.74^{+0.16}_{-0.17}$ & 5377 $\pm$ 31          & 2.62 $\pm$ 0.06        & $-190.59^{+2.74}_{-2.26}$ & $-2.17^{+0.28}_{-0.43}$ \\
 $19$ & $840727860939569152$             & $-12.50469$ & $-0.23993$ & $15.14$ & $5252^{+47}_{-54}$  & $2.35^{+0.18}_{-0.30}$ & 5291 $\pm$ 25          & 2.42 $\pm$ 0.05        & $-189.35^{+2.75}_{-2.07}$ & $-2.22^{+0.25}_{-0.50}$ \\
 $20$ & $1573344450075107456$            & $-10.91051$ & $-0.73733$ & $15.65$ & $5394^{+35}_{-44}$  & $2.50^{+0.16}_{-0.22}$ & 5382 $\pm$ 31          & 2.65 $\pm$ 0.05        & $-201.09^{+2.75}_{-2.89}$ & $-2.36^{+0.40}_{-0.49}$ \\
 $23$ & $1576400508285430272$            & $-4.18422$  & $-2.58857$ & $13.72$ & $5038^{+109}_{-3}$  & $1.02^{+0.97}_{-0.02}$ & 5000 $\pm$ 45          & 1.67 $\pm$ 0.05        & $-233.72^{+0.80}_{-0.69}$ & $-1.99^{+0.05}_{-0.04}$ \\
  \hline
      &                                  &             &            &         &                     & ISIS Non-members       &                        &                        &                           &                         \\
\hline
$1$   & $684655586991056512$             & $-59.94303$ & $1.28547$  & $12.84$ & $5062^{+2}_{-2}$    & $2.99^{+0.01}_{-0.02}$ & 4890 $\pm$ 55          & 1.52 $\pm$ 0.06        & $49.77^{+0.50}_{-0.45}$   & $-1.20^{+0.09}_{-0.09}$ \\
$2$   & $685227642275283456$             & $-57.91505$ & $0.80182$  & $13.39$ & $5064^{+3}_{-4}$    & $2.89^{+0.04}_{-0.03}$ & 4825 $\pm$ 25          & 1.76 $\pm$ 0.06        & $94.09^{+0.68}_{-0.65}$   & $-1.35^{+0.11}_{-0.20}$ \\
$11$  & $747560017309217792$             & $-39.49137$ & $-1.97128$ & $13.07$ & $5040^{+3}_{-5}$    & $2.64^{+0.03}_{-0.04}$ & 4857 $\pm$ 40          & 1.62 $\pm$ 0.06        & $-4.39^{+0.60}_{-0.61}$   & $-1.53^{+0.08}_{-0.08}$ \\
\hline
      &                                  &             &            &         &                     & LAMOST DR7             &                        &                        &                           &                         \\
\hline
--    & $793009399891751040$             & $-45.78102$ & $0.08422$  & $15.92$ & --                  & --                     & 5487 $\pm$ 36          & 2.97 $\pm$ 0.04        & 23.45$\pm8.56$            & -2.19$\pm0.10$          \\
--    & $793910484031755264^*$           & $-44.65316$ & $0.09744$  & $15.62$ & --                  & --                     & 5388 $\pm$ 60          & 2.82 $\pm$ 0.04        & 15.11 $\pm7.85$           & -2.23$\pm0.10$          \\
--    & $802849960243599872$             & $-37.86740$ & $0.43362$  & $15.88$ & --                  & --                     & 5490 $\pm$ 37          & 2.93 $\pm$ 0.05        & -27.40 $\pm7.14$          & -2.21$\pm0.10$          \\
 --   & $829645264771612800$             & $-27.34517$ & $0.17882$  & $15.99$ & --                  & --                     & 5316 $\pm$ 50          & 2.89 $\pm$ 0.06        & -102.87$\pm6.36$          & -1.79$\pm0.05$          \\
 --   & $832022718144536192^*$           & $-22.00954$ & $-0.03175$ & $15.97$ & --                  & --                     & 5346 $\pm$ 62          & 2.85 $\pm$ 0.06        & -142.68$\pm6.86$          & -2.08$\pm0.13$          \\
 --   & $1573762367572774272$            & $-10.41675$ & $-0.24961$ & $15.74$ & --                  & --                     & 5002 $\pm$ 46          & 1.67 $\pm$ 0.05        & -207.65$\pm4.63$          & -2.30$\pm0.03$          \\

 \hline 
\end{tabular} 
\end{table*}

Using the bright high-probability RGB sample defined above we obtained spectra
using the Intermediate-dispersion Spectrograph and Imaging System (ISIS) at the
4.2m William Herschel Telescope (WHT; program \texttt{SW2019a05}). We use the
R300B configuration for the blue-arm, yielding a R$\sim$300 in the wavelength
range of [325, 520] nm. The red-arm was setup in the R1200R mode, yielding a
R$\sim$5000 in the Calcium Triplet spectral region ([830, 890] nm). The
exposure time was adjusted such as to yield a S/N of at least 10 at 338nm in
the blue-arm. 

We targeted 23 stars based on the membership selection outlined in Section
\ref{sec:memb}, these were observed in the course of two nights in December
2019. Standard long-slit reduction was performed using \texttt{IRAF},
wavelength calibration was done using CuThAr arc-exposures taken at after each
exposure and at the end and beginning of the night. Because of the
varying slit losses and the possibility of atmospheric dispersion affecting the
spectra towards the blue wavelength, no attempt was made to flux the spectra.

We note that at the time of the observations, only Gaia DR2 data was available,
and 23 targets were selected for observation, out of which only the 16 higher
priority stars were observed due to scheduling and weather. Coincidentally the
stars that were not observed turned out to not pass our selection criteria with
the updated Gaia eDR3 data. For completeness and compatibility with the
observing proposal data, we chose to keep the star ids running from 1 to 23,
even though only 16 stars were observed.

We derived the effective temperature and gravity for the stars in our sample
with the code
\textsc{Brutus}\footnote{\url{https://github.com/joshspeagle/brutus}}
\citep{Speagle21}, using as input the Pan-STARRS and 2MASS photometry and a
prior on the distance coming from the stream track (with 0.5 kpc
uncertainties). These values are reported in Table \ref{tab:summary}
and are the ones used in our analysis of the CN/CH spectral features as well
for the Carbon abundances (see Section \ref{sec:abundances}).

The radial velocities of the stars in our spectroscopic follow-up were
derived from the one-dimensional reduced and calibrated spectra using
\texttt{rvspecfit} \citep{Koposov:2011, rvspecfit1} with \textsc{PHOENIX v2.0}
spectral library \citep{Husser:2013}. To obtain reliable uncertainties, we use
a Markov-Chain Monte Carlo (MCMC) minimisation that allows for radial velocity,
as well as $T_{\rm eff}$, $\log g$, and metallicity to be free. We estimated
the best-fit values as the median of the posterior chains while taking the 25
and 75 percentiles as their respective uncertainties. The radial velocities and
stellar parameters derived from this procedure are reported in Table
\ref{tab:summary}. We note that the agreement between the spectroscopic and
photometric stellar parameters is very good, and the overall results and
conclusions presented below are not dependent on the choice between
spectroscopic or photometric $T_{\rm eff}$ and $\log g$.

Although the fitting process described above was applied to spectra from both
ISIS arms, we chose to use the red-arm for the velocity and metallicity, since
it is higher resolution and allows for a comparison with CaT metallicity
determinations. These values can also be found in Table \ref{tab:summary}.  In
Figure \ref{fig:isismemb} we show the method used to select members based on
the inferred radial velocity and metallicities. We find that for the 16 Gaia
eDR3 selected stars observed 13 are likely members based on a $\feh < -1.75$
selection alone and find a weighted average metallicity of $-2.06\pm0.10$ for
this metallicity selected sample. We compare our metallicities to those derived
using the CaT equivalent width \citep{Vasquez:2015} and find them to agree
within 0.1 dex. When comparing the offset in radial velocity with GD-1 best-fit
orbit, we observe that all low-metallicity stars fall within 2$\sigma$ ($\sim
7.5 \kms$) of the mean, with the outliers (marked as non-members) being the
three highest metallicity stars.  We thus conclude that a metallicity selection
is sufficient to weed out contaminants in our ISIS sample, as it has already
been demonstrated Figure \ref{fig:cugivr}. The LAMOST DR7 sample was originally
constructed using our narrow PM+CMD selection and with added radial velocities,
we find that this selection is enough to weed out most contaminants. A single
high-metallicity star was identified in this sample and removed from the
analysis. We note that the range in $\pone$ where the three ISIS non-members
are located is also where GD-1's orbital line of sight velocity is the most
similar to field stars.

In Figure \ref{fig:spatial} we show the spatial distribution high-probability
RGB members. Confirmed RGB members from \citet{Li:2018} are shown, as well as
all the stars where ISIS or LAMOST DR7 spectra is available, however only the
former has labels corresponding to their ID (see Tab. \ref{tab:summary}). We
limit the spectroscopic samples to the magnitude range of the RGB. We also
allow for a fainter selection (black dots) which illustrates the location of
the bulk of GD-1's stars.

We find two members in the \emph{spur} region, star \#12 which is reported by
\citet{Li:2018}, and \#8 closer to the main stream track. We also find stars
\#16, \#17, and \#23 to be in the region associated the
\citep[blob/cocoon;][]{PriceWhelan:2018, Malhan:2018}. Stars \#17, and \#23 are
the most track deviant GD-1 confirmed members ever detected. We note that very
few RGB candidates remain to be followed up in the main body of GD-1, possibly
indicating that most RGB belonging to this stream have been found already.

\subsection{CN, CH, and Carbon abundances}
\label{sec:abundances}

\begin{table}
\caption{Measured CH and CN indices, derived carbon abundances and corrections \citep{Placco:2014}.}
\label{tab:abun}
\setlength{\tabcolsep}{4pt}
\begin{tabular}{lcccc}
\hline 
ID       &               CN & CH               &            A(C) & $\Delta$A(C)$_{{\rm corr}}$ \\
-        &              mag & mag              &             dex & dex           \\

\hline
\multicolumn{5}{c}{ISIS}\\
\hline

 $4$     &  $-0.13 \pm 0.03$ & $-0.42 \pm 0.04$ &$6.67 \pm 0.09$ & $0.01$    \\
 $6$     &  $-0.19 \pm 0.02$ & $-0.43 \pm 0.03$ &$6.52 \pm 0.08$ & $0.01$    \\
 $8$     &  $-0.19 \pm 0.02$ & $-0.45 \pm 0.04$ &$6.42 \pm 0.13$ & $0.01$    \\
 $9$     &  $-0.17 \pm 0.02$ & $-0.47 \pm 0.02$ &$5.39 \pm 0.07$ & $0.13$    \\
 $10$    &  $-0.20 \pm 0.02$ & $-0.43 \pm 0.03$ &$6.69 \pm 0.11$ & $0.01$    \\
 $12$    &  $-0.16 \pm 0.02$ & $-0.41 \pm 0.02$ &$6.09 \pm 0.11$ & $0.40$    \\
 $13$    &  $-0.18 \pm 0.02$ & $-0.38 \pm 0.03$ &$6.47 \pm 0.10$ & $0.01$    \\
 $16$    &  $-0.18 \pm 0.02$ & $-0.42 \pm 0.03$ &$6.43 \pm 0.08$ & $0.01$    \\
 $17$    &  $-0.18 \pm 0.02$ & $-0.43 \pm 0.03$ &$6.76 \pm 0.12$ & $0.01$    \\
 $18$    &  $-0.19 \pm 0.03$ & $-0.44 \pm 0.04$ &$6.74 \pm 0.09$ & $0.01$    \\
 $19$    &  $-0.19 \pm 0.04$ & $-0.42 \pm 0.04$ &$6.64 \pm 0.09$ & $0.01$    \\
 $20$    &  $-0.16 \pm 0.03$ & $-0.45 \pm 0.04$ &$6.50 \pm 0.06$ & $0.01$    \\
 $23$    &  $-0.16 \pm 0.02$ & $-0.40 \pm 0.03$ &$6.39 \pm 0.08$ & $0.33$    \\
\hline
\multicolumn{5}{c}{LAMOST DR7}\\
\hline
 696138$^{a}$  &	$-0.03 \pm 0.10$ & $ -0.40 \pm 0.07$ & $6.62 \pm 0.13$ &  0.01  \\
 793009 	 &	$-0.13 \pm 0.06$ &  $-0.45 \pm 0.03$ & $6.91 \pm 0.13$ &  0.01  \\
 793910 	 &	$-0.12 \pm 0.04$ &  $-0.44 \pm 0.04$ & $6.73 \pm 0.15$ &  0.01  \\
 802849 	 &	$-0.15 \pm 0.05$ &  $-0.44 \pm 0.05$ & $6.76 \pm 0.13$ &  0.01  \\
 829645 	 &	$-0.15 \pm 0.04$ &  $-0.46 \pm 0.03$ & $6.65 \pm 0.19$ &  0.01  \\
 832022 	 &	$-0.20 \pm 0.06$ &  $-0.34 \pm 0.04$ & $7.14 \pm 0.16$ &  0.01  \\
 832080$^{b}$ &	$-0.12 \pm 0.04$ &  $-0.44 \pm 0.02$ & $6.46 \pm 0.14$ &  0.01  \\
 157376 	 &	$-0.12 \pm 0.04$ &  $-0.43 \pm 0.02$ & $6.57 \pm 0.15$ &  0.01  \\
 157640$^{c}$  &	$-0.10 \pm 0.03$ &  $-0.42 \pm 0.03$ & $6.43 \pm 0.14$ &  0.33  \\

 \hline 

\end{tabular} 
 \footnotesize{$^a$ star no. 6 in the ISIS sample, $^b$ star no. 13 in the ISIS sample,
 $^c$ star no. 23 in the ISIS sample.}
\end{table}

We have calculated the band strengths S$\lambda$3883 (CN) and CH$\lambda$4300
(CH) to investigate the presence of multiple populations in GD-1. This
technique has been extensively used for GCs in the Galaxy
\citep[e.g.][]{Kayser:2008,Martell:2009,Pancino:2010,Lardo:2013} and clusters
in its dwarf satellites
\citep[e.g.][]{Hollyhead:2017,Hollyhead:2018,Hollyhead:2019,Martocchia:2021}.
Indices sensitive to absorption by the 4300\AA~CH and the 3883\AA~CN bands were
measured as described in \citet{Norris:1979} and
\citet{Norris:1981}\footnote{The blue-arm ISIS spectra extend to wavelengths
    where in principle the NH indices could be measured, however, the spectra is
of too low S/N at those wavelengths.}. Its is extremely difficult to establish
a continuum around the 3833\AA~CN band, where many atomic and molecular
absorption features are present. Thus, no attempt was made to normalise the
spectra before computing spectral indices
\citep[e.g.][]{Kayser:2008,Pancino:2010}. The 4300\AA~CH index measurement is
independent of this issue as a continuum can be established on both the blue
and red sides of the molecular absorption. Errors on measurements are
calculated assuming Poisson statistics, following \citet{Vollmann:2006}. The
derived uncertainties reflect the formal statistical uncertainties of the index
measurements. Systematic uncertainties are likely higher but they are not
relevant in this context since we are interested in relative differences
between band strengths.

\begin{figure}
    \centering
    \includegraphics[width=0.98\columnwidth]{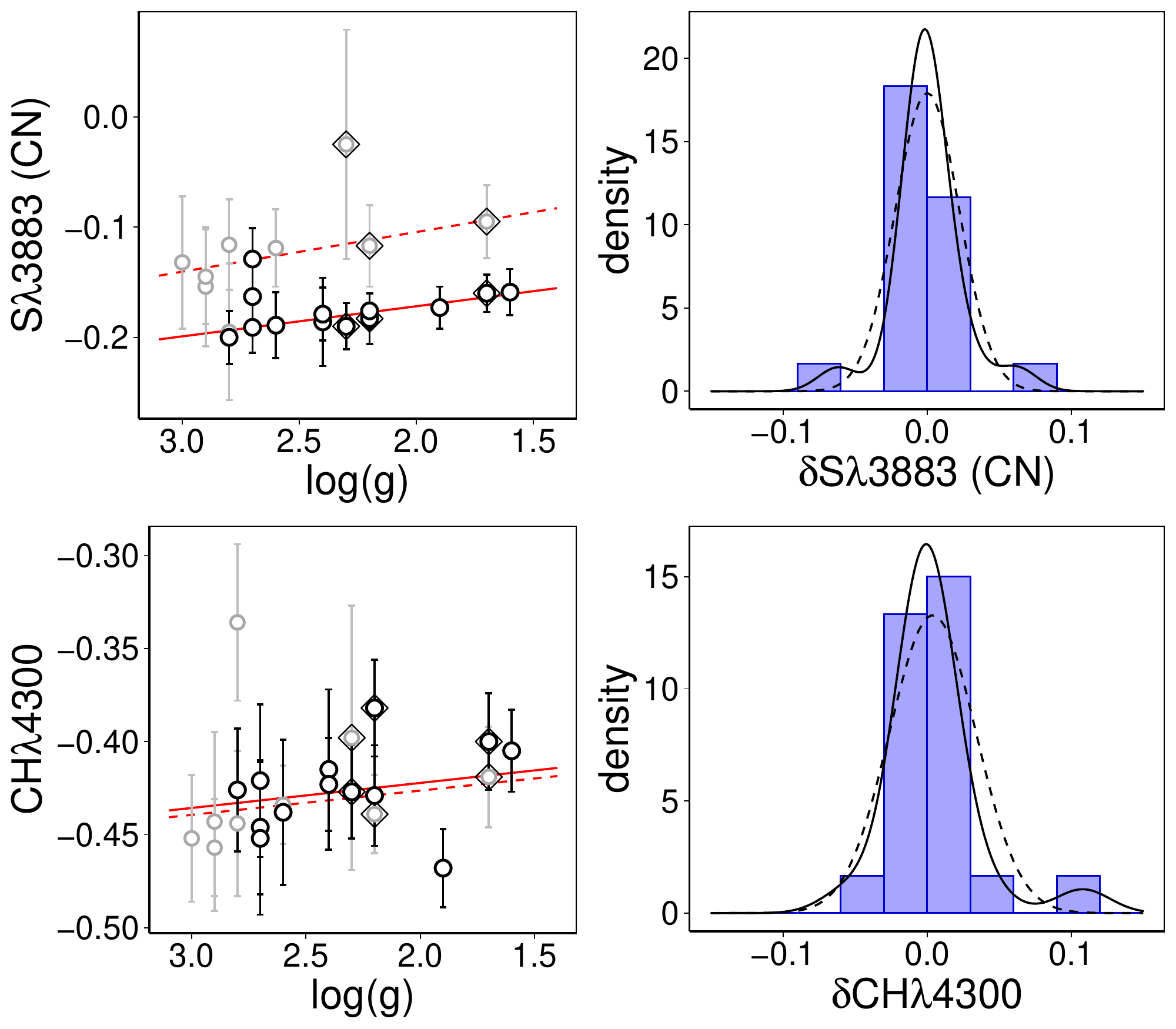}
    \caption{The left-hand panels show the run of the S$\lambda$3883(CN) ({\em
        top}) and CH$\lambda$4300 ({\em bottom}) indices against the surface
        gravity ($\log$ (g)) for GD1 member stars. Stars from the ISIS and
        LAMOST data-sets are plotted in black and grey colour, respectively.
        The solid and dashed red lines indicate the linear fit of those
        quantities versus magnitude for the ISIS and LAMOST spectra,
        respectively. Stars with available spectra from both ISIS and LAMOST
        are indicated with a large diamond. The right-hand panels show the
        histograms and the associated kernel distributions (solid black line)
        of the \dCN and \dCH ~residuals. The bandwidth of the Gaussian
        kernel density estimator was selected using the unbiased
        (least-squares) cross-validation bandwidth selector available in the
        \texttt{R} package \texttt{MASS} \citep{venables02}. The dashed line
        represents the Gaussian distribution that best fits the data.}
    \label{fig:abumag}
\end{figure}

In the left-hand panels of Figure~\ref{fig:abumag}, we show the strength of the
S$\lambda$3883 (CN) and CH$\lambda$4300 (CH) indices as a function of the
surface gravity ($\log$(g)) of each star in the ISIS and LAMOST sample (black
and grey symbols, respectively).  To account for any trend of the indices with
atmospheric parameters we fit a linear model by robust regression --which
attempts to down-weight the influence of outliers in order to provide a better
fit to the majority of the data \citep[e.g.,][]{venables02} -- between index
measurements and $\log$(g). This has been done for the ISIS and LAMOST
data-sets separately to properly take into account possible offsets between the
two data sources sets before analysing and interpreting results. Indeed, while
for the CH$\lambda$4300 index data from ISIS and LAMOST spectra are on the same
scale (see bottom left-hand panel of Figure~\ref{fig:abumag}), this is not the
case for the CN index. The top left-hand panel of Figure~\ref{fig:abumag} shows
that the $y$-intercept of the best fit linear model (b$_{0}$) is clearly
different for the two samples (b$_{0}$ =--0.117$\pm$0.013 and --0.033$\pm$0.044
for ISIS and LAMOST, respectively), whereas its slope is virtually identical.
This is likely because the ``one-sided'' CN index is more sensitive to
factors such as e.g. spectral resolution and quality of the flux calibration
than the  the ``two-sided'' CH index.  Differences between the measured
S$\lambda$3883 (CN) index from the ISIS and LAMOST spectra are of the order of
the $\Delta$S$\lambda$3883 (CN)$_{\rm (ISIS-LAMOST)}$= --0.65 mag for both ISIS
stars \#13 and \#23 (stars 832080 and 157640 in LAMOST, respectively), which is
comparable to the observed zero-point between the two data-sets. 
A larger difference is observed for ISIS star \#6 (star 696138 in LAMOST).
However the index measurements for this star are also characterised by very large
errors, so the observed discrepancy is likely due to the lower quality of the
LAMOST spectrum.

Indices corrected for temperature and gravity effects (denoted as \dCN~and
\dCH) have been obtained by subtracting the robust linear models shown in
Figure~\ref{fig:abumag} from the computed S$\lambda$3883 (CN) and
CH$\lambda$4300 indices. Their distribution is plotted in the right-hand panels
of Figure~\ref{fig:abumag}.  In the same panels the \dCN kernel density
distribution is also shown along with the best fit Gaussian\footnotemark. No evidence
for intrinsic \dCN~and~\dCH~variations that exceed measurement errors can be
derived from a careful inspection of the right-hand panels of
Figure~\ref{fig:abumag}. 

\footnotetext{To produce
histograms and kernel distributions for the {\em corrected} indices shown in
the right-hand panels of Figure~\ref{fig:abumag} we kept measurements  from the
ISIS data-set only for stars with spectra available in LAMOST.}

We further investigate the presence of any spreads by modelling the
distribution in~\dCN, and \dCH~ as a 1D normal distribution to our data. Here
we assume that for each data point a total dispersion can be computed in the
form of  $\sigma^2 = \sigma_0^2 + \sigma_j^2$, i.e. the sum in
quadrature of an intrinsic dispersion and the uncertainty in each $j$-th data
point.  We use this distribution to compute a likelihood that is maximized
using \texttt{emcee} \citep{Foreman-Mackey:2013}. We find and intrinsic
dispersion of $\sigma_{0, CH} = 0.011 \pm 0.008$ for \dCH; and  $\sigma_{0,
CN} = 0.008 \pm 0.007$ for \dCN. Both these dispersions are consistent with
zero at the 3-$\sigma$ level.  Thus, it is not possible to reveal any sign of
CN or CH intrinsic variations among GD1 stars from low-resolution spectra.

Although our ISIS spectra do not have the SNR to provide a reliable nitrogen
abundance, it was good enough to derive carbon abundances. These were inferred
by fitting observed spectra with synthetic ones in the spectral window from
4200 to 4400~\AA. 

To compute synthetic templates, we adopted the atmospheric parameters and
metallicities derived in Sec. \ref{sec:spec} and listed in Table
\ref{tab:summary}. Atomic and molecular line lists were taken from the most
recent Kurucz compilation from F. Castelli’s
website\footnote{\url{https://wwwuser.oats.inaf.it/castelli/}}. Model
atmospheres were calculated with the \texttt{ATLAS9} code \citep{Castelli:2004}
using the appropriate temperature and surface gravity for each star. We assumed
a microturbulent velocity v$_{\rm t}$= 2 km/s for all the stars. Kurucz's
\texttt{SYNTHE} code \citep{Kurucz:2005} was used to produce model spectra.
Observed spectra were normalised by using three continuum regions
(4200-4275\AA, 4315-4323\AA, and 4350-4440\aa) that are relatively free from
large molecular absorption (i.e. changes in carbon absolute abundances of 0.4
dex correspond in these spectral regions to flux variations that are less than
3\%). Finally, model spectra with varying carbon abundances were used in a
$\chi^2$ minimisation with the observed spectra to find the absolute carbon
abundances, A(C).

In order to calculate uncertainties from the fit parameters, we iteratively
change one parameter by its associated uncertainty and repeat the abundance
analysis. Finally, errors introduced by the fitting procedure were estimated by
re-fitting a sample of 100 spectra for each star after the introduction of
Poissonian noise in the best-fit template.

\begin{figure}
    \centering
    \includegraphics[width=0.37\textwidth]{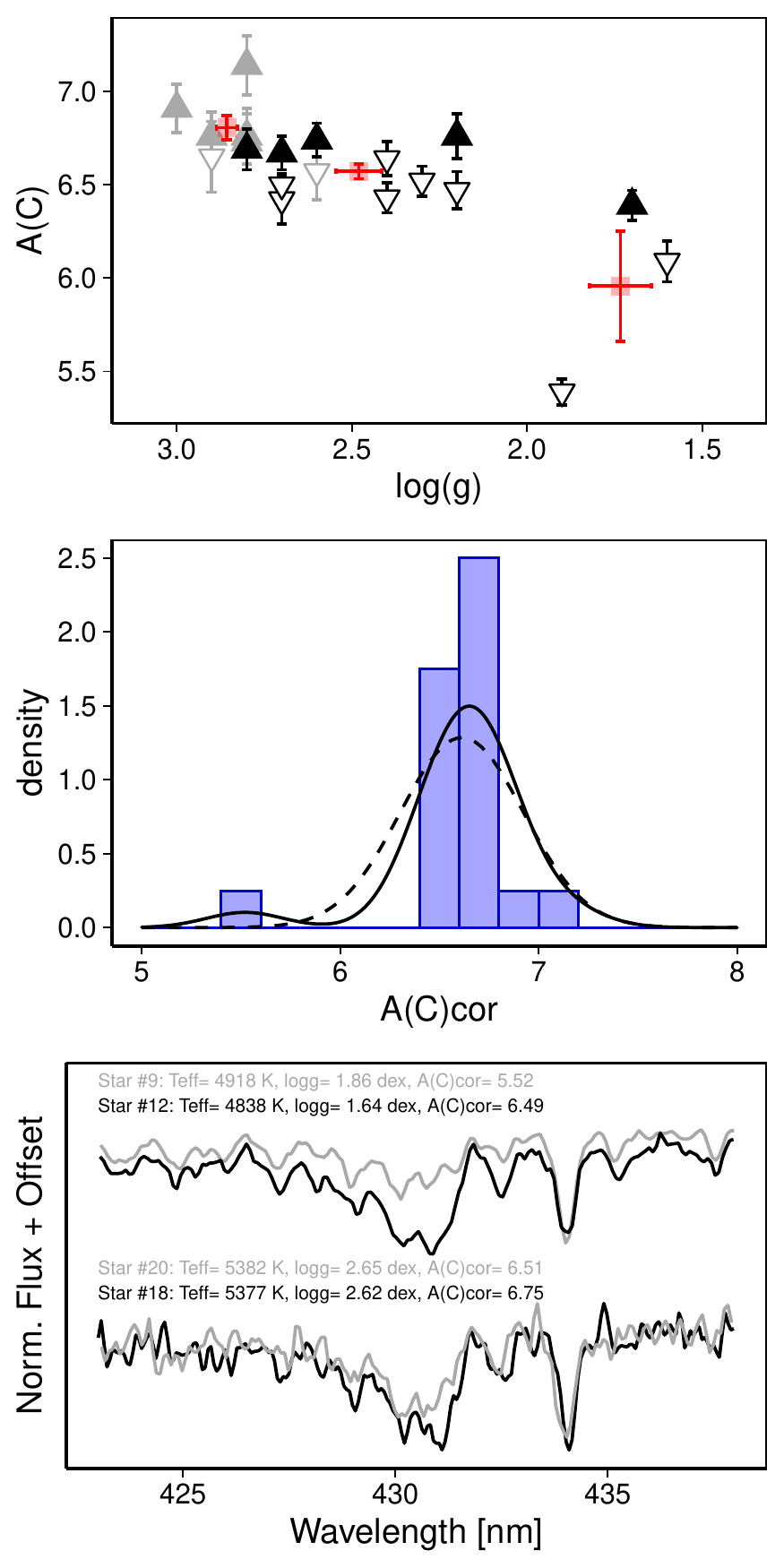}
    \caption{The top panels shows the measured absolute carbon abundances A(C)
        for all red giants in the sample, with stars identified as C-rich/poor
        with respect to the median A(C)cor value (see text)  shown as
        filled/empty symbols. The three large squares represent the mean and
        standard error in the luminosity bins: $\log (g) > $ 2.75, 2.0 $\leq
        \log (g) \leq $ 2.75, and  $\log (g) < $ 2.0. There is evidence for
        some deep mixing: stars in the brightest bin have on average lower
        carbon abundances than fainter stars.  The middle panel shows the
        histograms and the associated kernel distribution (solid black line) of
        the absolute carbon abundances corrected for evolutionary effects as in
        \citet{Placco:2014}.  The dashed line represents the Gaussian
        distribution that best fits the data. The bottom panel shows spectra of
        stars with similar atmospheric parameters yet very different C content
        (see legend). C-rich star spectra are in black, whereas the spectra of
        C-poor stars are plotted in grey.}
    \label{fig:cn_c}
\end{figure}

In the top panel of Figure~\ref{fig:cn_c} we show absolute carbon abundances
A(C) plotted against surface gravity. Increasing carbon depletion with rising
luminosity is observed in the data. This can be interpreted as a
sign of a mixing process that  brings partially processed CN material to
the stellar surface when stars evolve along the upper red-giant branch
\citep[e.g.][]{Sweigart79,Gratton00,Denissenkov03,Martell08b}. Thus, we used
the corrections of \citet{Placco:2014} to recover the initial carbon abundance
of our stars (i.e., carbon abundance not altered by internal stellar mixing).
According to \citet{Placco:2014}, carbon corrections are extremely small
(\mbox{$\Delta$A(C)} $\leq$ 0.02) for the majority of the stars analysed here.
Larger corrections (spanning values between \mbox{$\Delta$A(C)} = 0.13-0.40)
are only expected for the three brightest stars in the sample. We list the
index measurements, carbon abundances along with their associated uncertainties
for all targets in Table~\ref{tab:abun}. In the same table we also list a
evolutionary phase correction for carbon using the results from
\citet{Placco:2014}.

The distribution of the carbon abundances corrected for evolutionary mixing
effects, A(C)cor, is shown in the middle panel of Figure~\ref{fig:cn_c} along
with its associated kernel distribution and the Gaussian distribution which
best fits the data. While most of the stars have A(C)cor values around the median
value A(C)cor=6.67 with a small dispersion, there is a clear outlier (star \#9)
with A(C)cor=5.52$\pm$0.07. In the bottom panel of Figure \ref{fig:cn_c} we
show the spectrum of star \#9 around the CH absorption at 4300\AA~as well as
the spectrum of a star with similar atmospheric parameters but different C
abundance (star \#12, with A(C)cor=6.49$\pm$0.11). The CH band of star \#9
looks very weak compared to the stronger absorption observed in star \#12. We
also show in the same panel the spectra of two fainter objects (star \#18 and
\#20) with nearly identical stellar parameters yet different C content
(A(C)cor=6.75$\pm$0.09 and A(C)cor=6.51$\pm$0.06 for star \#18 and \#20,
respectively). A difference in the CH absorption region is observed also in
this case. Visual inspection of the bottom panel of Figure~\ref{fig:cn_c}
suggests that the intrinsic variations in carbon abundances may be present in
the analysed sample. We define C-rich/poor with respect to the median A(C)cor
value and use different symbols to plot C-rich and C-poor stars in the top
panel of Figure~\ref{fig:cn_c}.

In order to quantitatively assess the presence of an intrinsic spread in
carbon, we also repeated our 1-D normal distribution model fit to the corrected
A(C) measurements. This time we find that an intrinsic dispersion of
$\sigma_{0,C} = 0.32 \pm 0.06$. Given that star \#9 is an outlier, we also
repeat the fit without it, and find $\sigma_{0,C} = 0.13 \pm 0.04$. This points
towards some intrinsic dispersion in the C-abundance of GD-1 stars at a $\sim
3\sigma$ level.

To summarise, we found no statistically significant spreads in the CN and CH
indices among the observed RGB stars of GD-1. However, we find evidence for a
significant  C-abundance spread. 

Finally, we note that we found no obvious evidence for a spatial segregation of
stars with high/low C-abundance. This is supported by a 2-sample Kolmogorov–Smirnov
(KS) test, which
suggested that the C-poor and rich populations are drawn from the same
distribution in $\phi_1$ and $\phi_2 - f(\phi_1)$. This would indicate that
GD-1's progenitor was fully mixed by the time it started forming the stream.
Although a deeper and more homogeneous coverage of the stream would be
desirable to investigate this further.

\section{Apogee abundances}

When we crossmatched our sample of bona fide GD-1 RGB candidates with
the DR17 catalogue of APOGEE \citep{Majewski:17,DR17} we found two GD-1 stars
in common, stars \#4 and \#9 (both of these have also blue spectra, see Table
\ref{tab:summary}). Overall the stellar parameters and velocities inferred by
APOGEE are consistent with our independent measurements described in the
previous sections. Although APOGEE reports C, N, O and Al abundances for these
stars (most of the time with small uncertainties), a visual inspection of the
APOGEE spectra revealed that even the strongest spectral features that are
sensitive to these elements were weak and seriously affected by the noise.
Other studies have also cautioned about the reliability of the abundances
derived from APOGEE spectra in this temperature/metallicity regime
\citep[e.g.][]{Nataf19,Meszaros20}, due to this we will skip these elements in
the following discussion.

\begin{figure*}
    \centering
    \includegraphics[width=0.95\textwidth]{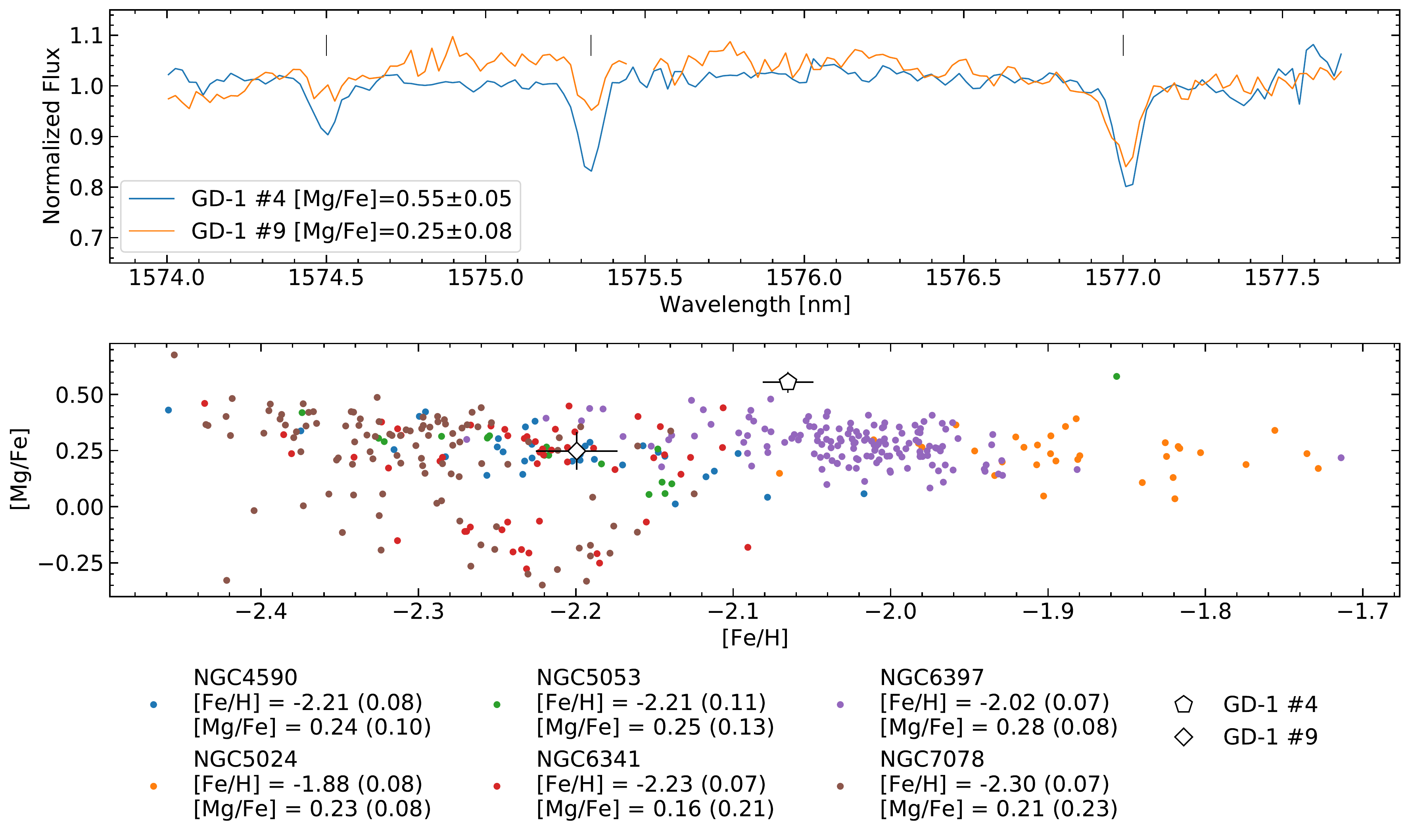}
    \caption{\emph{Top:} APOGEE DR17 spectra of  two GD-1 stars around strong
    Mg lines. \emph{Bottom:} Comparison between the APOGEE [Mg/Fe] of these
GD-1 stars and Galactic globular cluster stars of similar metallicity. For each
cluster, we quote the mean value of [Fe/H] and [Mg/Fe] and their associated
dispersion (standard deviation) in brackets.}
    \label{fig:apogee}
\end{figure*}

Unlike the elements mentioned above, the features sensitive to Mg stand
out from the noise in the APOGEE spectra, e.g.  top panel of Figure
\ref{fig:apogee}. For Mg, APOGEE reports a significant $\sim3.2\sigma$
difference between the [Mg/Fe] abundances of stars \#4 and \#9, see  Table
\ref{tab:apo}. In particular, star \#9, i.e. the star with the lowest
C-abundance (see the previous section), is also the one with the lowest [Mg/Fe] as
expected in P2 stars in globular clusters
\citep[see][]{Charbonnel16,BL18,Gratton19}.

\begin{table}
\caption{APOGEE parameters.}
\label{tab:apo}
\begin{center}
\resizebox{\columnwidth}{!}{
\begin{tabular}{lccccc}
\hline 
ID & $T_{eff}$ & $\log g$ & $V_{los}$ & [Fe/H] & [Mg/Fe] \\
- & K & dex & km/s & dex& dex\\
\hline
4 & $5091\pm24$ & $2.68\pm0.07$ & $98.7\pm0.7$ & $-2.07\pm0.02$  & $0.55\pm0.05$ \\
9 & $5068\pm47$ & $1.96\pm0.10$ & $-12.2\pm0.3$ & $-2.20\pm0.03$ &  $0.25\pm0.08$ \\
\hline
\end{tabular}} 
\end{center}
\end{table} 

For reference, in the bottom panel of Figure \ref{fig:apogee} we
compare the Mg-abundances of our GD-1 stars with the ones reported by APOGEE
for Galactic GCs of similar metallicities. The average $\mbox{[Mg/Fe]}\sim0.4
\mbox{~ dex}$ of these two GD-1 stars is slightly higher than the Galactic
clusters, and the difference in abundance between these two GD-1 stars is
$0.31\pm0.10 \mbox{~dex}$, which is comparable with the scatter observed in
Galactic globular clusters. Future studies with larger samples of stars, should
be able to quantify this behaviour more robustly.

In summary, the average [Mg/Fe] reported by APOGEE for these GD-1 stars
is slightly higher than the one found in Galactic GCs of similar metallicity.
Unfortunately, APOGEE only observed two of our GD-1 members, however, the
difference in [Mg/Fe] between them is larger than the one expected from the
reported uncertainties, and spans a similar range to the one observed in
globular cluster stars.

\section{Discussion and conclusion}

We demonstrate that a selection method using a combination of optical and
(near) UV colours are very successful in identifying RGB members of GD-1,
selected according to Gaia eDR3 PM.  In this particular case, we have an 81\%
success rate, which we attribute to the relatively low metallicity of GD-1
rendering its RGB colour bluer than the bulk of the field population. 

We are able to find members associated with off-stream features (\emph{spur}
and \emph{blob/cocoon}). We highlight stars \#23 and \#17, which lie 1.9\degree
and 2.6\degree away from GD-1 respectively. These two stars are the furthest
any confirmed GD-1 member has ever been found. Albeit being far from the stream
we find no significant velocity offset between the literature GD-1 orbit and
these off-stream stars. There are several scenarios for the formation of such
features \citep[e.g.][]{Erkal:2016, Bonaca:2018, Malhan:2018}, these all
predict a very small velocity offset, which is in agreement with what we observe
here. 

The study of Galactic and M31 globular clusters has revealed that the more
massive clusters show stronger signs of abundance variations than low-mass
clusters \citep[e.g.][]{Monelli13,Schiavon13}. This is likely a consequence
that for the most massive clusters the fraction of stars with anomalous
abundance is larger \emph{and} the magnitude of the abundance variations is
stronger \citep[see][]{Milone:2017}. The scaling relation between the way
the MPs manifest and the mass of a GC could serve as an independent proxy for
stream progenitor masses in the future.

Similarly, it has also been shown that for metal-poor GCs, smaller variations
in CN are expected as a consequence of the inefficiency to form this molecule
\citep{Martell:2008}. Both effects combined, i.e. a low $\sim10^4 {\rm ~\msun}$
mass and low metallicity ${\rm[Fe/H]}\sim -2.1 {\rm~dex}$, forecast that any
manifestations of the MPs in GD-1 are likely to be subtle.

In our analysis of optical spectra, we find no evidence for a significant
spread in CN or CH with the current (large) uncertainties.  However, we do find
evidence for an intrinsic spread in C-abundances, this detection is significant
(at the $\sim99\%$ level) even when we remove star \#9, which seems to be an
outlier in A(C). We observed no evidence for spatial segregation of stars with
different C-abundances in our sample, however, future studies with a more
homogeneous coverage of the stream should be able to address this more
robustly.

The lack of detectable nitrogen variations in stars with an intrinsic spread in
carbon is not surprising.  First, double-metal molecules like CN are
particularly difficult to observe at the overall low metallicity of GD-1
because their absorption strengths rapidly decline with decreasing metallicity
and their formation may be partially inhibited even in the presence of
substantial [C/Fe] and [N/Fe] differences \citep[e.g.][]{Sneden:74,Langer:92}.
Secondly, their formation depends both on carbon and nitrogen abundances. Deep
mixing scenarios predict that mixing efficiency should be high at
low metallicities \citep[e.g.][]{Sweigart:79}. Thus, evolved stars in a
low-metallicity cluster like GD-1 will be also depleted in carbon and enhanced
in nitrogen because of the dredge-up of CNO processed material. If nitrogen
becomes more abundant than carbon as a result of this process, the CN band
strength no longer scales monotonically with N abundance and can even decline
with rising luminosity \citep[e.g.][]{Smith:86,Martell:2008,Lee:21}.  As a
consequence, in the low-metallicity regime, the CN index has only limited
sensitivity to variations in nitrogen \citep[see Fig.~9 of][]{Martell:2008}. On
the contrary, absorption features of single-metal molecules like CH do not
weaken as much with decreasing metallicity, thus they still trace carbon
variations at metallicities [Fe/H] $<$ --2.0. 

We find that two of our confirmed GD-1 members are in the APOGEE DR17 sample.
From these two stars, we found that the mean [Mg/Fe] ($\sim0.4 \mbox{~dex}$) is
slightly higher than the one found in Galactic GCs of similar metallicities.
However, the GD-1 stars show a similar range in [Mg/Fe]  ($\sim 0.3
\mbox{~dex}$) than stars of metal-poor GCs. Moreover, the star with the lowest
[Mg/Fe] is also the star for which we inferred the lowest C-abundance (star
\#9), in agreement with the peculiar chemical patterns found in GC stars. This
suggests that GD-1 also manifests the multiple stellar population phenomenon
characteristics of more massive, undissolved globular clusters.

If the MPs signal is confirmed, it would make GD-1 another low mass GC
in the [Fe/H] < --2.0 regime to have displayed the phenomena
\citep{Li:2021, Ji:2020a}. This has implications both for the
different scenarios for MPs and for the origin of GD-1. In some scenarios
\citep[e.g.][]{Elmegreen:2017,Gieles:2018}, the formation of MPs is closely
tied to a high gas density/pressure environment, thus not allowing for such a
low initial mass system as GD-1 to have been formed in those conditions. One
could argue that GD-1's initial mass -- as inferred by \citet{Webb:2019} or
\citet{deBoer:2020} -- depends on the mass-loss while orbiting the Galaxy. It
is possible that GD-1 belonged to an accreted host galaxy, and had some of its
mass striped while in the original host. In this case, it could place it in an
initial mass range compatible with Galactic GCs. To confirm the accreted
scenario, observing the remnant of the host would be crucial to date the time
of accretion and host mass, allowing for stronger constraints on GD-1's initial
mass.

Even though Gaia helped immensely in the target selection for streams, looking for
additional evidence of light element spreads and for the possible host galaxy
that brought GD-1 into the Galaxy will be a task for future large high
multiplexing spectroscopic surveys. In the near future, we expect WEAVE
\citep{WEAVE}, to observe all of GD-1's RGB members in high/medium resolution,
and provide insight into MPs in this stream and their spatial distribution.
Using the same target selection technique used here, we expect to target up to
100 GD-1 members per linear degree in WEAVE's low-resolution halo survey.  This
work is the first study of the MPs phenomena for a completely dissolved GC and
we show the potential for using streams to open up a new region of parameters
space, not available in the local Universe population of GCs. 

We also point out, that for more metal-rich streams, where the CN band is
stronger, may be promising targets since their RGBs will split into multiple
components in $\cugi$ space, allowing for the study of the distribution of
different populations along the stream using photometry alone.

Finally, we would like to mention that our findings support the idea that the
small number (a few percent) of halo field stars known to display chemical
patterns characteristic of globular cluster stars originated from
disrupted/disrupting globular cluster
\citep[e.g.][]{Martell:16,Schiavon:17,Koch:19,Hanke:20,Horta:21}. 

\section*{Acknowledgements}

The authors would like to thank Nate Bastian, Mark Gieles, Amina Helmi, Jeremy
Webb, and Ricardo Schiavon for the comments and discussions during the early
stages of this work. We also thank Tadafumi Matsuno for the insight regarding
the APOGEE data and the anonymous referee for the comments and suggestions that
helped improve this work.

EB acknowledges support from a Vici grant from the Netherlands Organization for
Scientific Research (NWO).

Support for this work was provided by NASA through Hubble Fellowship grant
HST-HF2-51387.001-A awarded by the Space Telescope Science Institute, which is
operated by the Association of Universities for Research in Astronomy, Inc.,
for NASA, under contract NAS5-26555. This study was supported by the Klaus
Tschira Foundation.

CL acknowledges funding from Ministero dell’Università e della Ricerca through
the Programme ``Rita Levi Montalcini" (grant PGR18YRML1).

The William Herschel Telescope  and its service programme are operated on the
island of La Palma by the Isaac Newton Group of Telescopes in the Spanish
Observatorio del Roque de los Muchachos of the Instituto de Astrofísica de
Canarias.

This work has made use of data from the European Space Agency (ESA) mission
{\it Gaia} (\url{https://www.cosmos.esa.int/gaia}), processed by the {\it Gaia}
Data Processing and Analysis Consortium (DPAC,
\url{https://www.cosmos.esa.int/web/gaia/dpac/consortium}). Funding for the
DPAC has been provided by national institutions, in particular the institutions
participating in the {\it Gaia} Multilateral Agreement.

Funding for SDSS-III has been provided by the Alfred P. Sloan Foundation, the
Participating Institutions, the National Science Foundation, and the U.S.
Department of Energy Office of Science. The SDSS-III web site is
http://www.sdss3.org/.

SDSS-III is managed by the Astrophysical Research Consortium for the
Participating Institutions of the SDSS-III Collaboration including the
University of Arizona, the Brazilian Participation Group, Brookhaven National
Laboratory, Carnegie Mellon University, University of Florida, the French
Participation Group, the German Participation Group, Harvard University, the
Instituto de Astrofisica de Canarias, the Michigan State/Notre Dame/JINA
Participation Group, Johns Hopkins University, Lawrence Berkeley National
Laboratory, Max Planck Institute for Astrophysics, Max Planck Institute for
Extraterrestrial Physics, New Mexico State University, New York University,
Ohio State University, Pennsylvania State University, University of Portsmouth,
Princeton University, the Spanish Participation Group, University of Tokyo,
University of Utah, Vanderbilt University, University of Virginia, University
of Washington, and Yale University.

Guoshoujing Telescope (the Large Sky Area Multi-Object Fiber Spectroscopic
Telescope LAMOST) is a National Major Scientific Project built by the Chinese
Academy of Sciences. Funding for the project has been provided by the National
Development and Reform Commission. LAMOST is operated and managed by the
National Astronomical Observatories, Chinese Academy of Sciences. This paper
made use of the Whole Sky Database (wsdb) created by Sergey Koposov and
maintained at the Institute of Astronomy, Cambridge by Sergey Koposov, Vasily
Belokurov and Wyn Evans with financial support from the Science \& Technology
Facilities Council (STFC) and the European Research Council (ERC).

The following software packages where used in this publication:
    \package{Astropy} \citep{astropy, astropy:2018},
    \package{gala} \citep{gala},
    \package{IPython} \citep{ipython},
    \package{matplotlib} \citep{mpl},
    \package{numpy} \citep{numpy},
    \package{scipy} \citep{scipy}
    \package{vaex} \citep{Breddels:2018}

\section*{Data Availability}
All the survey data used in the work is publicly available. The reduced data
underlying this work is available from the authors upon reasonable request.

\bibliographystyle{mnras}
\bibliography{refs_full} 

\bsp	
\label{lastpage}
\end{document}